\documentclass [preprint,showpacs,prl]{revtex4}
\usepackage{graphicx}
\usepackage{bm}
\begin{document}
\narrowtext

\title{Enhanced hard X-ray emission from femtosecond laser irradiated microdroplets}
\author{M. Anand, A. S. Sandhu,  S.Kahaly, G. Ravindra Kumar and M. Krishnamurthy{\footnote {email:mkrism@tifr.res.in}}}
\address{Tata Institute of Fundamental Research, 1 Homi Bhabha Road, Mumbai 400 005, India.}
\author{P.Gibbon}
\address{John-von-Neumann Institute for Computing, ZAM, Forschungszentrum J\"{u}lich, D-52425 J\"{u}lich, Germany 
}

\date{\today}

\begin{abstract}
We make a comparative study of hard x-ray emission from 15 $\mu$m methanol microdroplets and a plain slab target of similar atomic composition at similar laser intensities. The hard X-ray yield from  droplet
plasmas is $\simeq$ 35 times more than that obtained from solid plasmas.  A
prepulse that is about 10ns and about 5\% of the main pulse is essential for
hard x-ray generation from the droplets.
 A hot electron temperature of 36 keV
is measured from the droplets at 8$\times$10$^{14}$ W cm$^{-2}$; three times
higher intensity is needed to obtain similar hot electron temperature from
solid plasmas with similar composition.  We use 1D PIC simulation to obtain
qualitative correlation to the experimental observations. 
\end{abstract}

\pacs{52.38.Ph, 52.25.Os, 79.20.Ds, 52.50.Jm}

\maketitle

        The physics of laser-plasma interactions has undergone a revolution in
recent times. Technological advances in lasers have opened the possibility of
achieving intensities up to 10$^{21}$ W cm$^{-2}$ \cite{21expts}. The
non-perturbative physics of laser-matter interactions at these extreme
intensities has brought forth many new concepts and applications
\cite{concepts}.  The hot dense plasma produced in such interaction has opened up novel schemes of pulsed neutron generation \cite{neutron},
nuclear reactions \cite{nuclear}table top
acceleration \cite{accelerate} and synchrotron radiation \cite{hhg}. Importantly, such plasmas are  
promising sources of ultrashort pulse radiation in EUV and x-ray regimes and   increasing the  
efficiency of these sources is a major challenge.  This obviously leads to the 
 investigation of strategies to efficiently couple laser energy to the plasma. One such strategy has been the introduction of novel
 targets . Metallic nanoparticle-coated solids \cite{rajeev} and `velvet'
 targets \cite{velvet} have yielded enhanced x-ray emission in the moderate to
 very hard x-ray regime. Gaseous clusters, which are nanoparticles with solid-like local
 density have been shown to absorb 90$\%$ of the incident laser energy
\cite{krainov}. Particle acceleration up to an MeV and efficient nuclear fusion at intensities as low as 10$^{16}$ W cm$^{-2}$ has been observed from such cluster plasmas.

There are however some disadvantages in the use of gaseous clusters. A
major one is the rather stringent limitation on  the type  of atomic or
molecular species which can  produce large clusters. For example, there is no
simple way to generate large clusters with high-Z atoms like Pt. Besides, there is very little hard X-ray emission above 5 keV from clusters
\cite{rhodes}. In this context, we invite attention to liquid droplets as a promising alternative. They are relatively
debris-less and couple the advantage of size confinement with the
relative ease with which a droplet can  be used to introduce any
atomic/molecular system of interest.  Droplet targets have found application
in EUV Lithography and X-ray microscopy \cite{EUVmicro,phillips}.  The
emphasis on droplet plasma studies has so far been mostly towards optimizing the EUV radiation  at 13 nm due to applications towards
lithography, though there have been some initial studies on hard x-ray radiation \cite{droplet-xray,milchberg,zhangAPB}. The
study of very hard X-ray emission from droplets is certainly of major interest.

In this paper, we present measurements of hard x-ray emission (10-350 keV), from 15$\mu$m methanol droplets irradiated with 100 fs laser
pulses with intensities up to 2$\times$10$^{15}$ W cm$^{-2}$.  We find that a
prepulse that arrives at about 10ns ahead of the main pulse is critically
important to generate hard x-rays from liquid droplets at these
intensities. For comparison, we measure hard X-ray emission from a solid
plastic target (which has a similar atomic composition) under similar conditions.  
Though the prepulse brings about 17 fold enhancement in the x-ray yield at
3.7$\times$10$^{15}$ W cm$^{-2}$ from the plastic , we find that the
size limited methanol droplet generates hard x-rays at
much lower intensity and  much more efficiently. The hard x-ray yield at 1.5$\times$10$^{15}$ W
cm$^{-2}$ is about 35  times larger than that from the plastic under similar
conditions. We present 1D PIC simulation that qualitatively explain the experimental observation of enhanced x-ray generation in microdroplets. 

The apparatus used in these experiments has been described elsewhere\cite{apb} and here we present only the salient features. 
The microdroplet targets are generated by forcing methanol through a 10 $\mu$m
capillary, which is modulated at 1 MHz using a piezo-crystal. The
uniformly sized droplets were characterized by imaging of the droplet,
and also by observing morphological dependent resonances (MDR)\cite{mdr}.  The
inset in figure 1(a) shows the droplets along with the image of a 25 $\mu$m
slit used for calibration.  The droplets are produced inside a vacuum chamber
maintained at 10$^{-5}$ Torr.  We focus the 100fs pulses of 800nm light using
a 30 cm planoconvex lens and achieve intensities up to 2 $\times$ 10$^{15}$ W
cm$^{-2}$. A two-pulse setup is used to obtain a prepulse at about 10ns
ahead of the main pulse and a pair of polarizers together with a half wave
plate is used to control its intensity. Comparative experiments with solids are performed on
an optically flat plastic by focusing p-polarized light at 45$^\circ$
incident angle with a 20 cm lens to a spot size of 20 $\mu$m and achieve
intensities up to 5$\times$ 10$^{15}$ W cm$^{-2}$. We have carefully
determined the  intensity of light by measuring the pulse width using a second
order autocorrelator (Femtochrome-103XL) and the beam waist (30$\mu$m)using
the standard knife edge technique. We have established the accuracy of this
method in the past by correlating the measured values with the well known
appearance intensity of Xe$^{q+}$ ions \cite{grk}. We used plastic targets for comparison 
since their atomic composition is close to  that of methanol. The
target is scanned such that each laser pulse is incident at a fresh portion of
the target \cite{mag-fields}. The x-ray detector in all experiments is a NaI(Tl) detector, appropriately time gated with the laser pulse and
calibrated with standard radioactive sources.

In experiments with liquid droplets, there is no measurable hard x-ray yield
 at intensities less than 1.5$\times$10$^{15}$W cm$^{-2}$ in the absence of a
 prepulse. In the regenerative amplifier, a prepulse can be generated by misalignment of the pockel cell. In initial experiments, we found that the hard x-ray generation was very sensitive to the extent of this prepulse, which is 10ns ahead of the main pulse.  Once we established that a ns
 prepulse is essential for the hard x-ray generation from the droplet, we 
 set up a two-pulse experiment to introduce an intentional prepulse of desired
 intensity that arrives at the required time ahead of the main pulse. We find
 that while a prepulse that is 1-10 ps ahead does not
 significantly influence the x-ray emission from the droplets, a prepulse that is about 10ns ahead
 is essential to produce x-rays from the droplets. The x-ray yield increases
 steeply with the prepulse energy and saturates for a prepulse
 that is about 5\% in intensity of the main pulse.

 The x-ray emission spectrum obtained from 15$\mu$m methanol droplets at a
 prepulse intensity of about 5\% is shown figure 1(a). The solid line shows an
 exponential fit to the data assuming a Maxwellian distribution for the
 electrons in the plasma. In this fit we only considered energies larger
 than 50  keV so that  corrections due to the transmission through the glass or
 aluminum housing of the detector are negligible. To avoid pile up, the count
 rate was kept less than 0.1 per pulse by restricting the solid angle of
 detection\cite{apb}. The X-ray emission spectrum from  the plastic at similar
 pre-pulse intensities is shown in figure 1(b) at about three times larger
 main pulse intensity, as there was no measurable X-ray emission below 2 $\times$10$^{15}$
 W cm$^{-2}$. Exponential fits to the data show that the hot electron
 temperature is about 40 keV for plastic while it is 36 keV in the case of methanol
 droplets at about three times less intensity.

A comparison between the relative integrated X-ray yields from both droplet
plasma and plastic target, with a prepulse of about 1.5$\times$ 10$^{14}$W
cm$^{-2}$, is shown in figure 2. The X-ray yields from both the targets are
measured in the range from 10-350 keV.  Experiments on liquid drops
with higher intensities are not possible with our present laser, as  we are
constrained to maintain a focal spot size of 30 $\mu$m  to maintain the
droplet close to the center of the focus,  given the spatial jitter  of a
few microns in the jet.  The prepulse enhances the x-ray
generation in plastic  but the enhancement from the methanol droplet  is
much larger. The threshold for hard x-ray generation in droplets is a factor of two
smaller,  and at an intensity of about 2$\times$10$^{15}$W cm$^{-2}$ the x-ray
yield from droplets is at least 35 times larger than that obtained from the plastic.

How do we model  the laser interaction with the droplet?
In plasmas made of mesoscopic matter, both the geometry and the size are crucially important. 
A microdroplet is a spherical cavity that interacts with light very
differently compared to a planar surface. On the
 droplet surface the angle of incidence would vary from 0$^\circ$ to
 90$^\circ$ and accordingly the polarization would change from s to p, as we
 go from the center to the poles of the drop. Also a microdroplet, much larger
than the wavelength of light, can focus the light
inside the drop. A  major fraction of the prepulse (10$^{13}$W cm$^{-2}$) is known to enter the droplet and very little is lost in ionization on the surface. The light that enters is focused by the droplet and its intensity is enhanced by  two orders of magnitude or more
\cite{white-nano}. For our droplet size, Lorentz-Mie  calculations \cite{bohren,matlab}
show that a maximum intensity enhancement of nearly 150 times the incident light intensity is possible at a few
spots inside  the droplet close to its surface (Figure 3). 

Focusing of the prepulse in a liquid drop  results in substantial ionization at many spots in and around the drop \cite{white-nano} and leads to a large volume  spherical plasma.   Imaging experiments using the pump-probe technique show that the droplet plasma is of  30$\mu$m in diameter, when the main pulse is incident\cite{apb}.  So, the main pulse is
incident on a large volume spherical plasma, that is close to the critical
density in case of a droplet target. 

Unraveling the dynamics of a spherical droplet plasma would require 3D PIC
 simulations, which are still too expensive to realistically model the present
 experimental conditions.  
However to gain useful insights into the differences between solid
 and droplet plasmas, we have carried out high-resolution 1D-PIC simulations
 with different density profiles  that qualitatively mimic the expected density
 profiles from  the droplet, at least in one dimension (see inset of
 Fig.4(a)). An upper limit for the plasma scale length created by the prepulse
 can be obtained from the isothermal model of Rosen
 \cite{rosen:spie90,bastiani:pre97}.  Assuming an absorbed flux of $2\times
 10^{13}$W cm$^{-2}$ for the prepulse, the plastic target would be initially heated
 to around 20~eV.  One-dimensional expansion at the sound speed would then
 give $L \sim c_s t \simeq 170 \mu$m after 10~ns.  Plasma cooling and geometrical factors
 will reduce this somewhat, but we can nevertheless expect density
 profiles with $L/\lambda > 10$.  The droplets will expand even more due to
 their limited mass and the Mie-enhancements in prepulse intensity. 

The simulations were performed using BOPS, a
 1D2V PIC code exploiting the Lorentz boost technique to handle
 oblique-incidence interactions \cite{gibbon:prl92,gibbon:pop99}.
  The  unperturbed solid plasma profile is represented by a 6$\mu$m plasma slab with steep sides
 ($L/\lambda <0.02$).  For the intensities used here ($I<5\times 10^{15}$
 W cm$^{-2}$) this was thick enough to prevent multiple reflection (and
 therefore additional heating) of hot electrons from the rear side of the target. For the droplet an exponential density
 ramp was included on both sides with $L/\lambda$ and the maximum density
 varied such that the total charge was the
 same as for the unexpanded slab.  At these
 low intensities and long plasma lengths, a large number of particles
 (typically 20 million electrons and ions) were
 needed in order to generate a statistically significant number of hot
 electrons above 20 keV. 

Although the simulated hot electron temperatures found in Fig.~4a) are roughly
 a factor of 2
 lower than those observed in the experiments, we find that the hot electron yield
 does show a qualitative correlation with the observations.  
 Fig. 4(b) shows the variation in  the hot  electron numbers  with density
 scale length, showing a 5-fold increase in electrons above 20 keV and an 
onset of `superhot' electrons ($>50$ keV) as the
 profile is stretched from an abrupt step to an extended corona. 
The general increase in hot electron temperature and number is due to the
 better matching of incidence angle (here fixed at 15 degrees) to the
 scale-length where resonance absorption is optimized.  In the experiment these conditions are
 better reproduced for the droplets than the solid targets, where the laser
was incident normally and at 45$^o$ respectively,
 reducing the resonant coupling to the plasma in the latter case.  At long scale-lengths
 ($L/\lambda>>1$) there may also be a significant contribution from parametric instabilities in the
 extended underdense region, which appear to be responsible for the very hot
 electrons observed.

One has to keep in mind
 that these calculations have been performed with one dimensional density profiles and are only
 qualitatively indicative of the experimental measurements. The effects could
 be larger if the sphericity of the target
 were included.

Not only is the generation of hot electrons different in a spherical droplet, but escape of plasma electrons away from the target is different depending on the target geometry.
Recently it  was shown that 
at
any given distance away from the plasma source, the hot electron fraction is
likely to be much larger from a spherical target compared to a plane solid
slab \cite{fill}. The experimental results presented here correlate well with this simple analytical model.

In summary, we have studied x-ray emission from  intense laser irradiation of 
15 $\mu$m methanol droplets in comparison with that from solids. Our results
show that the yields from  droplet plasmas are larger by  a factor of ~35. A
prepulse 10ns ahead of the main pulse is essential for efficient hard x-ray generation from the
droplets. The preplasma from a spherical droplet is arguably more extensive
than from the plane slab target and is conducive to efficient hot electron
generation via resonance absorption. This idea is supported by 1D PIC
simulations that mimic the long scale-length droplet profile. 

We thank D. Mathur,  E. Springate, M. J. Vrakking, and J. Jha  for suggestions,help and advice. Support from the DST(GOI) for our laser system is acknowledged.

\begin{figure}
\vskip 7in
\special{eps: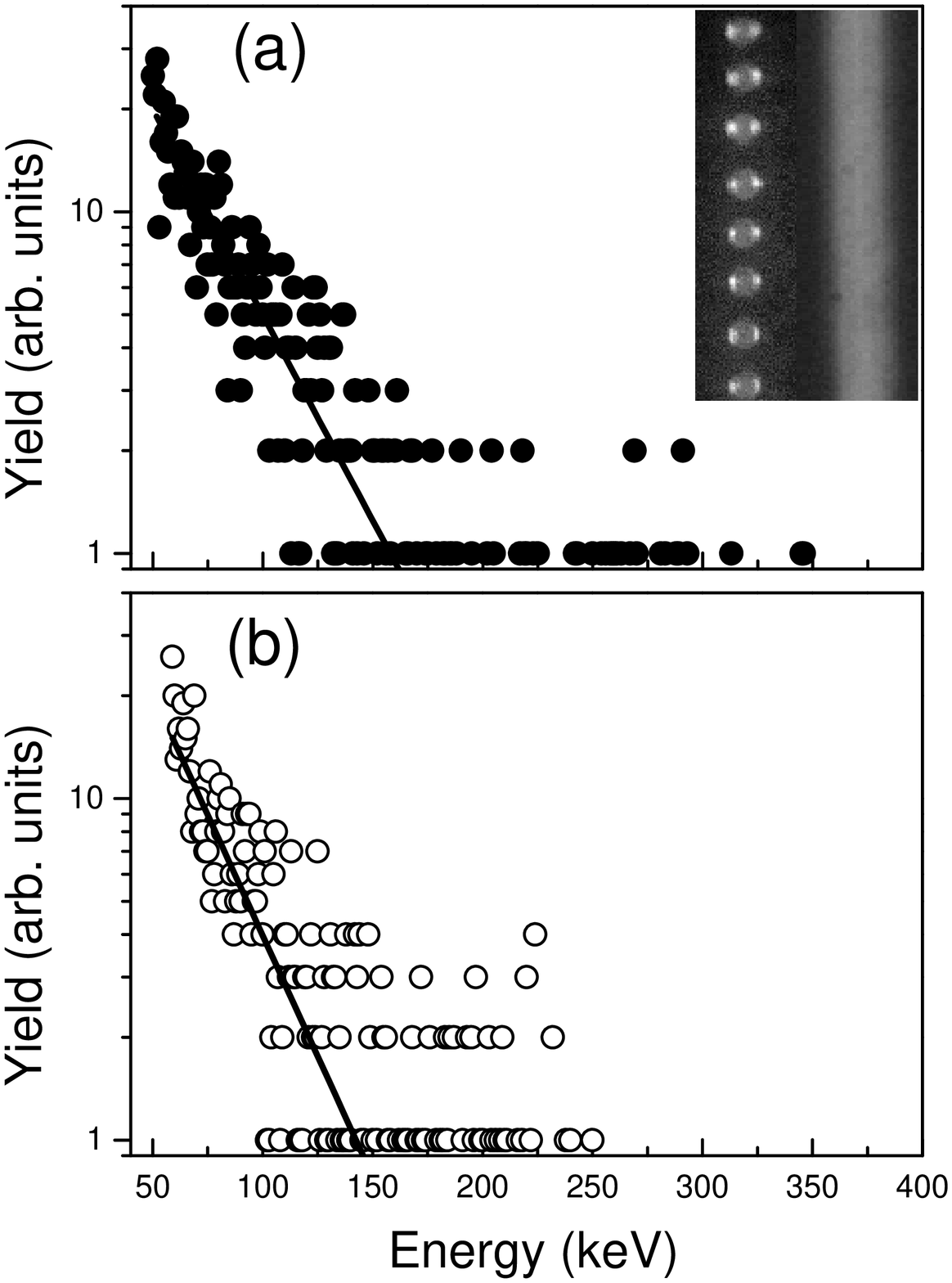 x=6in y=7in }
\caption{a). X-ray emission spectra obtained when 15$\mu$m methanol droplet are irradiated at 8$\times$10$^{14}$ W cm$^{-2}$. The solid line shows the least square fit to the data assuming a Maxwellian distribution for the electrons of 36 keV temperature. The inset shows an image of the droplet stream along with the image of a precision 25 $\mu$m slit used for determining the size of the droplet. b) X-ray emission spectra obtained when a solid plastic target is irradiated with similar laser pulses at an intensities of 2 $\times$10$^{15}$ W cm$^{-2}$ at 45$^\circ$ to the normal using P-polarized light. The solid line shows the least square fit to the data assuming a Maxwellian  distribution for electron of 40 keV temperature.
 }
\end{figure}

\begin{figure}
\vskip 7in
\special{eps: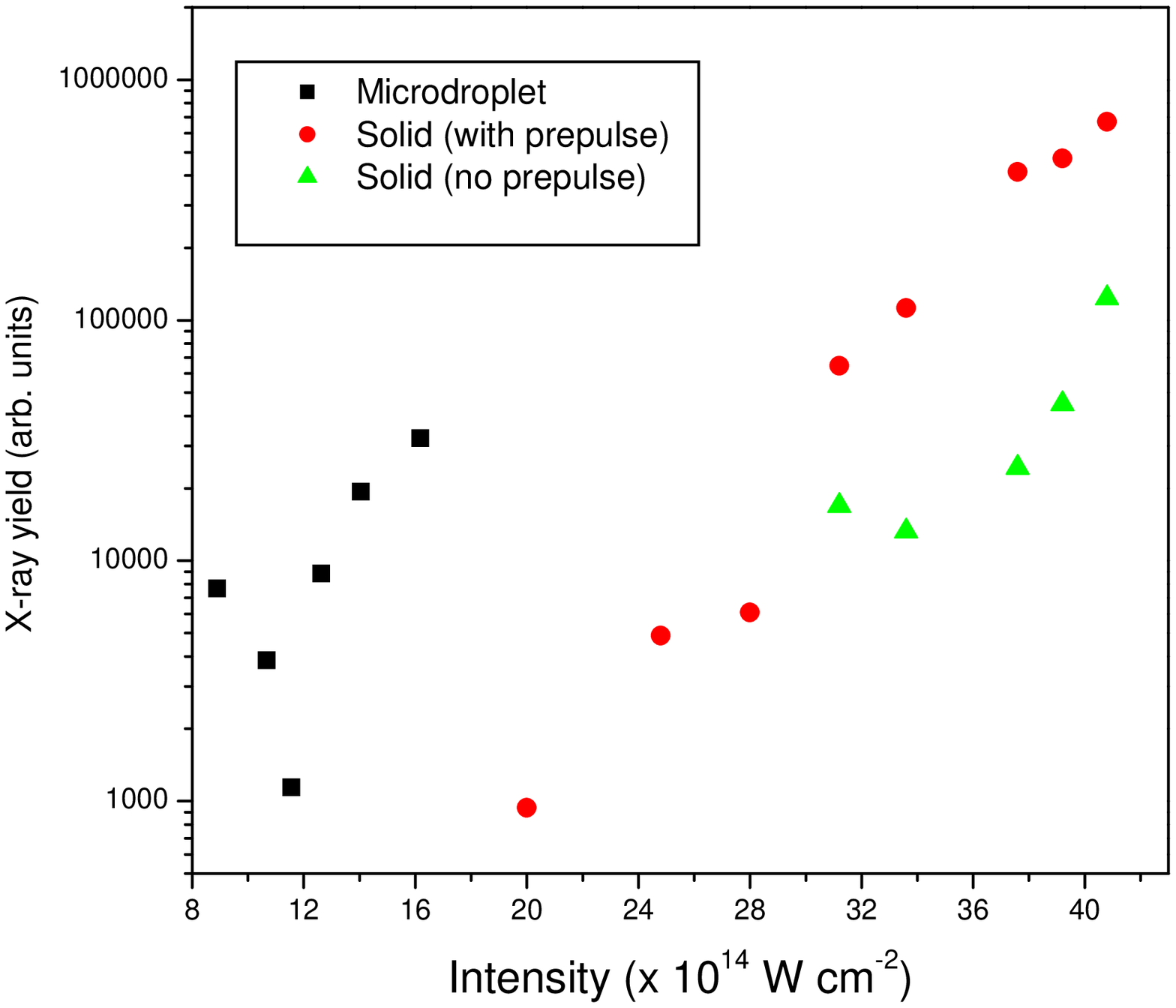 x=6in y=7in }
\caption{ X-ray emission yields measured for 15$\mu$m methanol liquid droplet targets (squares) and solid plastic target (circles) as a function of the incident intensities and pulse energies. The x-ray yield from solid target without prepulse is also shown(triangles).}
\end{figure}

\begin{figure}
\vskip 7in
\special{eps: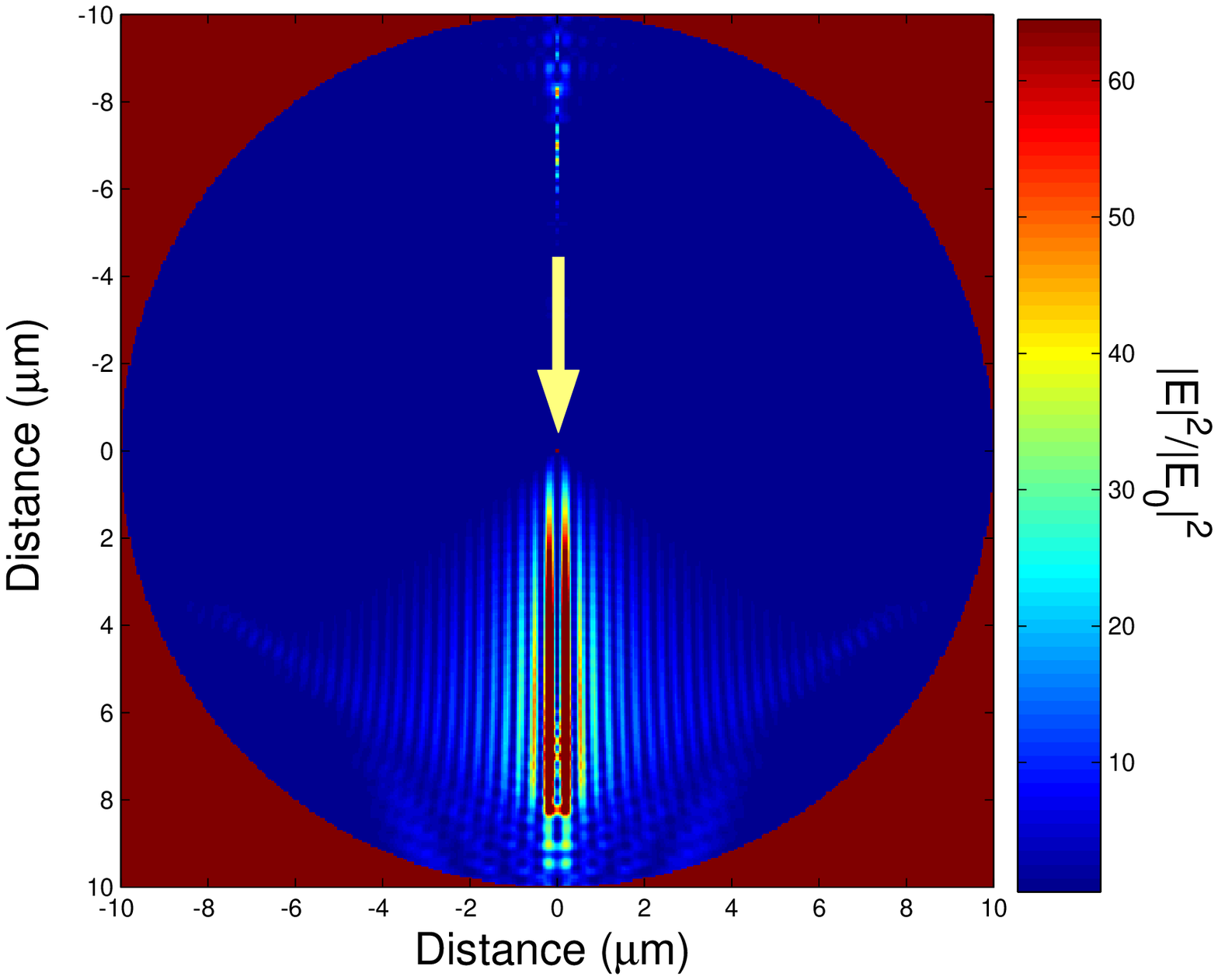 x=6in y=8in }
\caption{Computed  ratio of the internal to incident absolute square electric field inside the droplet cavity calculated for a plane wave incident on a droplet using the Lorentz-Mie theory. The arrow indicates the direction of laser propagation.}
\end{figure}

\begin{figure}
\vskip 7in
\special{eps: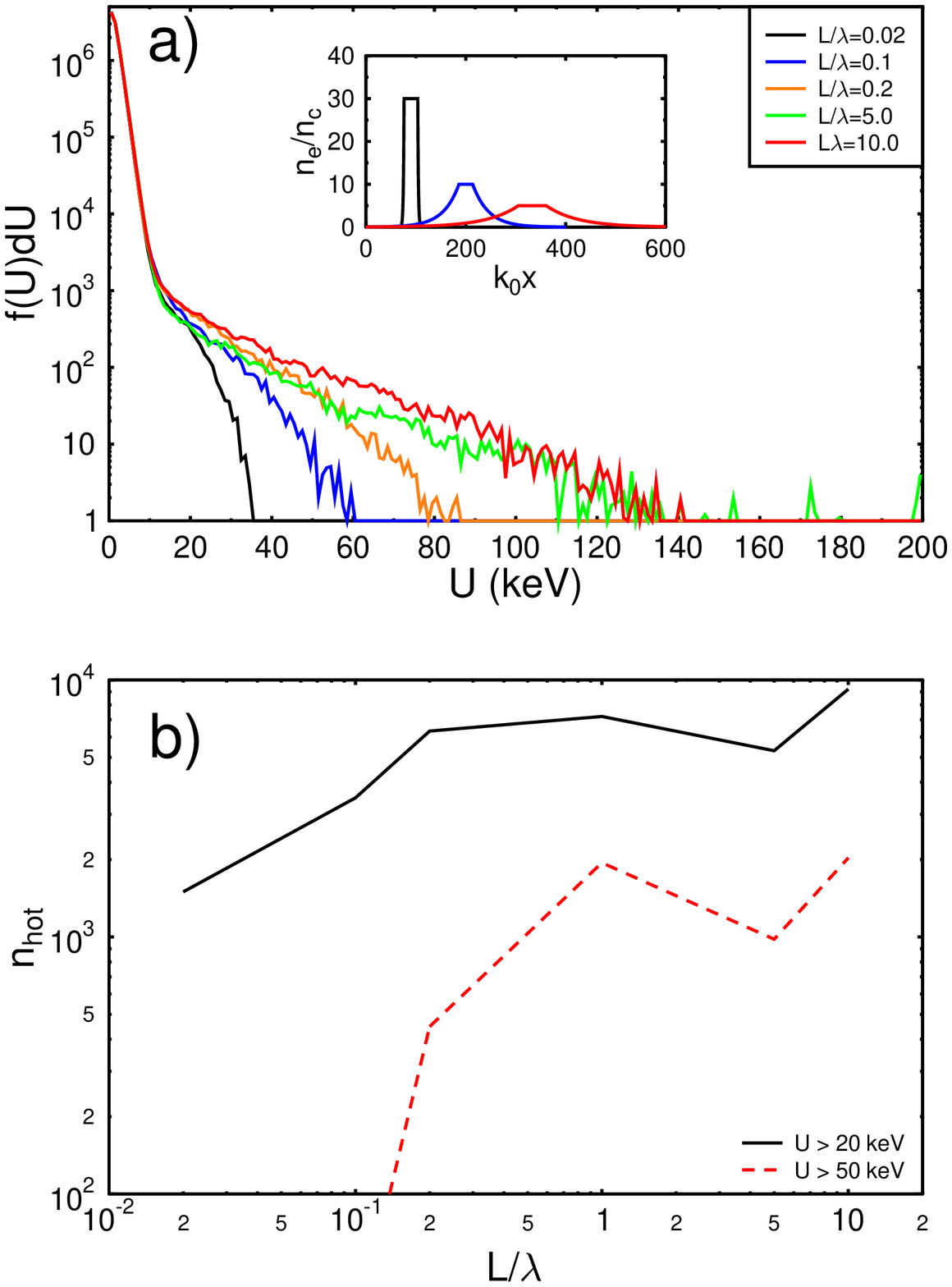 x=6in y=7in }
\caption{a)Electron energy distributions calculated using 1D PIC simulation
  with the density profiles that mimic the droplet (inset) 
b) Hot electron numbers extracted from spectra in a) as a function of density
gradient. The lines are drawn to guide the eye.}
\end{figure}

\end{document}